%% file: main.tex
\documentclass{article}
\usepackage{amsmath,graphicx,mlspconf}
\usepackage{multirow}
\usepackage[table,xcdraw]{xcolor}
\usepackage{amsmath,amssymb,amsfonts,dsfont}
\usepackage{stmaryrd}
\usepackage{url}
\usepackage{dsfont}
\include{notations}

\makeatletter
\def\bstctlcite{\@ifnextchar[{\@bstctlcite}{\@bstctlcite[@auxout]}}
\def\@bstctlcite[#1]#2{\@bsphack
  \@for\@citeb:=#2\do{%
    \edef\@citeb{\expandafter\@firstofone\@citeb}%
    \if@filesw\immediate\write\csname #1\endcsname{\string\citation{\@citeb}}\fi}%
  \@esphack}
\makeatother

\usepackage[textwidth=2.0cm, textsize=tiny]{todonotes}
\title{A deep learning architecture to detect events\\in EEG signals during sleep}
\newcommand*\samethanks[1][\value{footnote}]{\footnotemark[#1]}

\name{Stanislas Chambon$^{1, 2, 3}$\sthanks{Contributed equally. 
This work was supported in part by the french Association Nationale de la Recherche et de la Technologie (ANRT) under Grant 2015 / 1005}, Valentin Thorey$^{2}$\samethanks[1], Pierrick J. Arnal$^{2}$, Emmanuel Mignot$^{1}$, Alexandre Gramfort$^{3, 4, 5}$}
\address{
$^1$ Center for Sleep Sciences and Medicine, Stanford University, Stanford, California, USA\\
$^2$ Research \& Algorithms Team, Dreem, Paris, France\\
$^3$ LTCI T\'{e}l\'{e}com ParisTech, Universit\'{e} Paris-Saclay, Paris, France\\
$^4$ Inria, Universit\'{e} Paris-Saclay, Paris, France\\
$^5$ CEA Neurospin, Universit\'{e} Paris-Saclay, Paris, France\\
}

\graphicspath{{figures/}}

\begin{document}
\bstctlcite{IEEEexample:BSTcontrol}
%

\maketitle
\begin{abstract}

Electroencephalography (EEG) during sleep is used by clinicians to evaluate various neurological disorders. In sleep medicine, it is relevant to detect macro-events ($\geq 10$\,s) such as sleep stages, and micro-events ($\leq 2$\,s) such as spindles and K-complexes. Annotations of such events require a trained sleep expert, a time consuming and tedious process with a large inter-scorer variability. Automatic algorithms have been developed to detect various types of events but these are event-specific. We propose a deep learning method that jointly predicts locations, durations and types of events in EEG time series. It relies on a convolutional neural network that builds a feature representation from raw EEG signals. Numerical experiments demonstrate efficiency of this new approach on various event detection tasks compared to current state-of-the-art, event specific, algorithms.
\end{abstract}

\input{1_introduction}
\input{2_method}
\input{3_experiments}
\input{4_discussion}
\input{5_conclusion}


\bibliographystyle{IEEEtran}
\bibliography{biblio}

\end{document}

%% file: notations.tex
\def \R {\mathbb{R}}
\def \E {\mathbb{E}}
\def \N {\mathbb{N}}
\def \X {\mathcal{X}}
\def \Y {\mathcal{Y}}

\def \F {\mathcal{F}}
\def \EE {\mathcal{E}}

\DeclareMathOperator*{\argmin}{arg\,min}
\DeclareMathOperator*{\argmax}{arg\,max}

\def\x{{\mathbf x}}
\def\L{{\cal L}}

%% file: 1_introduction.tex
\section{Introduction}
\label{sec:intro}

During sleep, brain activity is characterized by some specific electroencephalographic (EEG) patterns, or events, (e.g. spindles, K-complexes) used to define more global state (e.g. sleep stages)~\cite{Iber2007}. Detecting such events is meaningful to better understand sleep physiology~\cite{Warby2014,Purcell2017} and relevant to the physiopathology of some sleep disorders~\cite{Stephansen2017,Manoach2016,Musiek2015}. Traditionally, visual analysis of these events is conducted, but it is tedious, time consuming and requires a trained sleep expert. Agreement between experts is often low, although this can be improved by taking consensus of multiple sleep experts~\cite{Warby2014}.

Automatic detection algorithms have been proposed to detect specific types of micro-events in sleep EEG. These typically build on a band-pass filtering step within a certain fixed frequency band, for instance $11 - 16$\,Hz for the detection of spindles.  Three types of methods can be distinguished. The first type relies on extracting the envelop of the filtered signal and performing a thresholding-like step with either a fixed or tunable threshold~\cite{Ray2015, Wamsley2012, Wendt2012, Moelle2011, Nir2011, Ferrarelli2007}. It is primarily used for spindle detection. The main difference between these lies in the thresholding part which is either performed on the rectified filtered signal~\cite{Ferrarelli2007}, or on the instantaneous amplitude obtained after an Hilbert-Transform~\cite{Nir2011}, on the root mean square of the filtered signal~\cite{Moelle2011} or on the moving average of the rectified filtered signal~\cite{Wamsley2012}. This first type of methods can identify start and end times of events by looking at inflexion points of the envelop of the filtered signal~\cite{Wendt2012}. Furthermore, most of these approaches are specific to a sleep stage ~\cite{Ferrarelli2007, Nir2011, Moelle2011, Wamsley2012, Ray2015} and rely heavily on preprocessing steps such as notch filtering around 50\,Hz to remove the electrical current artefacts or general visual artefacts removal, impractical on large amounts of data. The second type of methods relies on decomposing the EEG into its oscillatory and transient components. Then filtering and thresholding are used to detect events of interest~\cite{Parekh2017, Lajnef2017, Parekh2015}. These are more general methods since they can detect either sleep spindles or K-complexes and can work over entire sleep recordings independently of sleep stages. The third type of methods corresponds to machine learning methods and are more general. These filter the EEG signal, extract spectral and temporal features from a segment and predict with a binary SVM whether it is a spindle~\cite{Lachner-Piza2018}. These methods all have significant limitations: (1) use of band-pass filtering with fixed cut-off frequencies that might not be adapted to some subjects (2) they are intrinsically event specific (3) their hyper parameters are often optimized on the record used for evaluating the performance, introducing an overfitting bias in reported results.


A solution to these problems may be found in the adaptation of recent computer vision algorithms developed in the context of object detection. Indeed, detecting events in multivariate EEG signals consist in predicting time locations, duration and types of events which is closely related to the object detection problem in computer vision where bounding boxes and objects categories are to be predicted. For this latter problem, state-of-the-art methods make use of dedicated deep neural networks architectures~\cite{Lin2017,Liu2016, Redmon2015,Ren2015}.

In this paper, we introduce such a dedicated neural network architecture to detect any type of event over the sleep EEG signal. The proposed approach builds on a convolutional neural network which extracts high-level features from the raw input signal(s) for multiple default event locations. A localization module predicts an adjustment on start and end times for those default event locations while a classification module predicts their labels. The whole network architecture is trained end-to-end by back-propagation. In this paper, we first present the general method and the architecture of the proposed neural network. We then evaluate performance of the proposed approach versus the current state-of-the-art methods on $2$ event detection tasks.

%% file: 2_method.tex
\paragraph*{Notation}
We denote by $\llbracket n \rrbracket$ the set $\{1, \ldots, n\}$ for $n \in \N$. Let $\X \in \R^{C \times T}$ be the set of input signals, $C$ is the number of EEG channels and $T$ the number of time steps. $\L = \llbracket | L | \rrbracket$ stands for labels of events where $L$ is the number of different labels. $0$ is the label associated to no event or background signal. An event $e = \left\{t^c, t^d, l \right\} \in \EE = \R^2 \times \L \cup \{0 \}$ is defined by a center location time $t^c$, a duration $t^d$ and an event label (or type) $\l \in \L$. A \emph{true event} is an event with label in $\L$ detected by a human scorer or a group of human scorers, \emph{a.k.a.} \emph{gold standard}. A \emph{predicted event} is an event with label in $\L$ detected by an algorithm or a group of algorithms.
%
%

\section{Method}

\begin{figure}[ht!]
\centering
\includegraphics[width=0.9\linewidth]{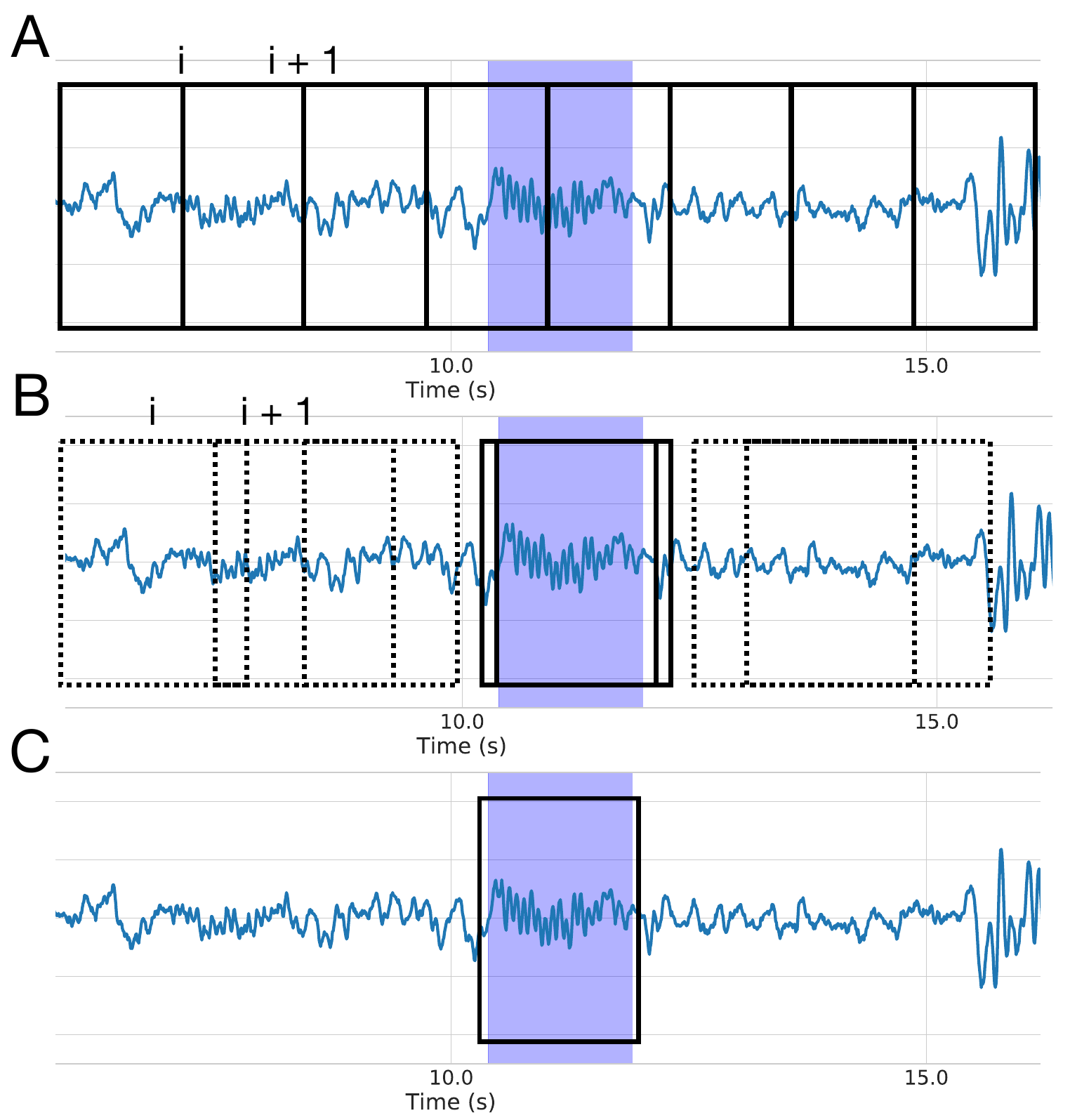}
\caption{\label{fig:scheme}Proposed approach prediction procedure inspired by SSD~\cite{Liu2016}: A: default events are generated over the input signals. B: the network predicts refined locations of any default event and its label included the label no event: $0$. C: non-suppression maximum is applied to merge overlapping events with label different from $0$. The network finally returns the locations of the merged events and their labels}
\end{figure}

The detection procedure employed by our predictive system is illustrated in Figure~\ref{fig:scheme} (with C=1).
Let $x \in \X$ be an input EEG signal. First, default events are generated over the input signal, \emph{e.g.} 1\,s events every 0.5\,s if this corresponds to a typical duration of events to be detected (cf. Figure~\ref{fig:scheme}-A). Positions, durations and overlaps of default events can be modified. Then, the network predicts for each default event its adjusted center and duration, together with the label of the potential event (cf. Figure~\ref{fig:scheme}-B). Events with the highest probabilities are selected, and non-maximum suppression is applied to remove overlapping predictions (cf. Figure~\ref{fig:scheme}-C). This is similar to the SSD~\cite{Liu2016} and YOLO approaches~\cite{Redmon2015} developed in the context of object detection. 
%
Training this predictive system requires the following steps. First, default events are generated over the input signal and matched to the true events based on their Jaccard index, \emph{a.k.a.} Intersection over Union (IoU)~\cite{Redmon2015}. The network is trained to predict the centers and durations together with the labels of the events. Default events which do not match a true event with a sufficient IoU are assigned the label  $l = 0$. To address the issue of label imbalance between real events and events with label  $l = 0 $, subsampling is used so that only a fraction of events with label $l = 0$ is used for training.

In order to learn a system to achieve the prediction task just described with back-propagation, one needs to design a fully differentiable architecture. The aim is to learn a function $\hat{f}$ from $\X$ to $\Y$ where $y \in \Y$ is a set of elements from $\EE$. Let $N_d$ be the number of default events generated over the input signal $x \in \X$. It is also the number of adjusted events predicted by the network. Let $D(x) = \left\{ d_i = (t^c_i, t^d_i), i \in \llbracket N_d \rrbracket \right\}$ be the set of centers and durations of the $N_d$ default events generated over $x$. Let $E(x) = \left\{ e_j = (t^c_j, t^d_j, l_j) : j \in \llbracket N_e \rrbracket \right\}$ be the list of the $N_e$ true events annotated over the signal $x$.
Default events which match a true event are selected to train the localization and classification capacities of the system. The $\mathrm{IoU}(d_i, e_j) \in [0,1]$ is computed between each default event $d_i \in D(x)$ and each true event from $e_j \in E(x)$. Let $\eta > 0$, $d_i$ matches $e_j$ if $\mathrm{IoU}(d_i, e_j) \geq \eta$. If multiple true events match the same default event $d_i$, the true event which exhibits the highest IoU with $d_i$ is selected. We introduce the function $\gamma$ which returns, if it exists, the index of the true event matching with the default event $d_i$, and $\emptyset$ otherwise:
\begin{equation*}
\gamma(i) = \argmax \limits_{\substack{ j \in \llbracket N_e \rrbracket \\ \mathrm{IoU}(d_i, e_j) \geq \eta}} \mathrm{IoU}(d_i, e_j) \in \llbracket N_e \rrbracket \cup \{\emptyset\}
\end{equation*}

Let $d_i$ be a default event matching with the true event $e_j$. $d_i$'s center and duration are then encoded with $\phi_{e_j}: \R^2 \longrightarrow \R^2, d_i = (t^c_i, t^d_i) \longmapsto \left( \dfrac{t^c_j - t^c_i}{t^d_i}, \log \dfrac{t^d_j}{t^d_i} \right)$~\cite{Ren2015}. This encoding function quantifies the relative variations in centers and durations between the default event $d_i$ and the true event $e_j$, and represents the quantities the network actually predicts.
%
%
Let $\hat{f}(x) \in \Y$ be the prediction made by model $\hat{f}$ over $x$. We define it as $\hat{f}(x) = \{(\hat{t}^c_i, \hat{t}^d_i, \hat{l}_i) \in \EE, i \in \llbracket N_d \rrbracket \}$. $\hat{\tau}_i = (\hat{t}^c_i, \hat{t}^d_i)$ are the predicted coordinates of encoded default event $d_i$ and $\hat{l}_i$ is its predicted label. In practice, the model will output the probability of each label $l \in \L \cup \{ 0 \}$ for default event $d_i$ so $\hat{l}_i$ is replaced by $\hat{\pi}_i \in [0, 1]^{ |\L| + 1}$. As it is a probability vector, we have $\sum_{l \in \L \cup  \{0 \}} \hat{\pi}_i^l = 1$.
%
%

The loss between the true annotation $E(x)$ and the model prediction $\hat{f}(x)$ over signal $x$ is
a function $\ell: \Y \times \Y \rightarrow \R_+$ defined as $\ell \left( E(x), \hat{f}(x) \right) = \ell^+_{norm} + \ell^-_{norm}$ where
\begin{align}
\ell^+ &= \sum_{\substack{ i \in \llbracket N_d \rrbracket \\ \gamma(i) \neq \emptyset }} \mathrm{L1}_{smooth} \left( \phi_{ e_{\gamma (i)} }(d_i), \hat{\tau}_i \right) - \log(\hat{\pi}_i^{l_{\gamma (i)}}) \label{eq:loc_clf_pos}\\
\ell^- &= - \sum_{\substack{i \in \llbracket N_d \rrbracket \\
                  \forall j \in \llbracket N_e \rrbracket: \mathrm{IoU}(d_i, e_j) < \eta}}
            \log(\hat{\pi}_i^0)\label{eq:clf_neg}
\end{align}

%
%
$\ell^{+}_{norm}$ (resp. $\ell^-_{norm}$) is obtained by dividing $\ell^{+}$ (resp. $\ell^-$) by the number of terms in the sum \eqref{eq:loc_clf_pos} (resp. \eqref{eq:clf_neg}). In \eqref{eq:loc_clf_pos}, we sum the localization and classification loss for any default event $d_i$ matching a true event $e_{\gamma(i)}$. The $\mathrm{L1}_{smooth}$ loss applies coordinate-wise the real valued function: $x \mapsto (x^2 / 2) \mathds{1}_{|x| < 1} + (|x| - 1/2) \mathds{1}_{|x| \geq 1}$~\cite{Ren2015}. Equation~\eqref{eq:clf_neg} stands for the classification loss of default boxes which do not match any true event. \color{black}Subsampling has been ommitted in \eqref{eq:clf_neg} for simplicity. In practice, we use a $1 / 3$ ratio between default events matching true events and those not matching a true event, selecting those with the worst classification scores.\color{black}

In the end, the learning problem consists in solving the following minimization problem to obtain event detector $\hat{f}$:
\begin{equation}\label{eq:training_objective}
\hat{f} \in \argmin_{f \in \F} \E_{x \in \X} \left[ \ell \left(E(x), f(x) \right) \right]
\end{equation}

In order to specify what is the function class $\F$, one needs to detail the network architecture. We consider a general convolutional network that, given a set of default events $D(x) = \{d_i = ( 2^8 \cdot ( i - 0.5) / \rho, \ t_i^d): i \in \llbracket N_d \rrbracket \}$, predicts $N_d$ events, where $N_d = T \cdot \rho / 2^8$ and $\rho \in \N$ is an overlapping factor. The network is composed of 3 parts, see Table~\ref{tab:network_architecture}.

The first part, called Block 0, performs a spatial filtering in order to increase the Signal to Noise Ratio (SNR) by recombining the original EEG channels into virtual channels using a 2D spatial convolution~\cite{Chambon2018}. It takes as input a tensor $x \in \X$ and outputs a new tensor $x_0 \in \X$. Channels of $x_0$ are obtained by linear combination of the channels of $x$. It can be seen as a 2D convolution with $C$ kernel of size (C, 1), a stride of $1$ and a linear activation. It is followed by a transposition to permute the channel and spatial dimensions in order to recover a tensor $x_0 \in \X$. If $C=1$, this block is skipped.

The second part, composed of Block $k$, for $k \in \llbracket 8 \rrbracket$, performs feature extraction over $x_0$ in the time domain. Each block is composed of a 2D convolution layer with batch normalization~\cite{Ioffe2015} and ReLU activation $x \mapsto \max(x, 0)$~\cite{Nair2010}, followed by a temporal max-pooling. Block $k$ first convolves the previous feature maps $x_{k-1}$ with $4 \times 2^k$ kernels of size $(1, 3)$ (space, time), using a stride of $1$. Zero padding is used to maintain the dimension of the tensor through the convolution layer. Then, the ReLU activation is applied. Finally a temporal max pooling operation with kernel of size $(1, 2)$ and stride $2$ is applied to divide by 2 the temporal dimension. Each Block $k$ does not process the spatial dimension.
%
%

The third part, Block 9 takes as input a tensor $x_8 \in \R^{1024 \times C \times T / 2^8}$, and for each default event $i \in \llbracket N_d \rrbracket $, it predicts the event label and its encoded center and duration. Block 9 has two layers: a classification layer 9-a and a localization layer 9-b. Layer 9-a convolves the last feature maps with $(|\L| + 1) \times \rho$ kernels of size $(C, 3)$ (stride of $1$). A \emph{softmax} operation is applied along the channel dimension on every set of $(|\L| + 1)$ channels so that, for each default event $d_i$, one obtains the probability $\hat{\pi}_i^l$ of the corresponding event $i$ to belong to any of the classes $l \in \L \cup \left\{ 0 \right\}$. Similarly, layer 9-b convolves the last feature maps $x_8$ with $2 \times \rho$ kernels of size $(C, 3)$ and stride $1$. This gives the predicted coordinates $\hat{\tau}_i = (\hat{t}_i^c, \hat{t}_i^d)$ of any encoded default event $i \in \llbracket N_d \rrbracket $.

\input{2a_network}

%% file: 2a_network.tex
\begin{table*}[ht!]
\small
\centering

\begin{tabular}{llllllll}
\rowcolor[HTML]{FFFFFF} 
                                                                                                                 & Layer          & Layer Type     & \# kernels       & output dimension                   & activation                                                          & kernel size & stride \\
\rowcolor[HTML]{F0F0F0} 
\cellcolor[HTML]{F0F0F0}                                                                                         & 1              & Convolution 2D & C                & (C, 1, T)                          & linear                                                              & (C, 1)      & 1      \\
\rowcolor[HTML]{F0F0F0} 
\multirow{-2}{*}{\cellcolor[HTML]{F0F0F0}Block 0}                                                                & 2              & Transpose      & -                & (1, C, T)                          & -                                                                   &             & -      \\
\rowcolor[HTML]{F0F0F0} 
                                                                                                                 &                &                &                  &                                    &                                                                     &             &        \\
\rowcolor[HTML]{F0F0F0} 
\cellcolor[HTML]{F0F0F0}                                                                                         & k-a          & Convolution 2D & $4 \times 2 ^ k$ & $(4 \times 2^k, C, T / 2^{k - 1})$ & ReLU                                                                & (1, 3)      & 1      \\
\rowcolor[HTML]{F0F0F0} 
\multirow{-2}{*}{\cellcolor[HTML]{F0F0F0}\begin{tabular}[c]{@{}l@{}}Block k\\ for $k \in \llbracket 8 \rrbracket$\end{tabular}} & k-b          & Max Pooling 2D & -                & $(4 \times 2^k, C, T / 2^k)$   & -                                                                   & (1, 2)      & 2      \\
\rowcolor[HTML]{F0F0F0} 
                                                                                                                 &                &                &                  &                                    &                                                                     &             &        \\
\rowcolor[HTML]{D2D0D0} 
\cellcolor[HTML]{D2D0D0}                                                                                         & 9-a & Convolution 2D & $(| \L | + 1) \times \rho$    & $( | \L | + 1) \times \rho, 1, T / 2^8)$        & \begin{tabular}[c]{@{}l@{}}softmax\\ (channel dimension)\end{tabular} & (C, 3)      & 1      \\
\rowcolor[HTML]{D2D0D0} 
\multirow{-2}{*}{\cellcolor[HTML]{D2D0D0}Block 9}                                                                & 9-b   & Convolution 2D & $2 \times \rho$              & $(2 \times \rho, 1, T / 2^8)$        & linear                                                              & (C, 3)      & 1     
\end{tabular}
\caption{Model architecture: Block 0 performs spatial filtering and outputs a tensor $\x_0 \in \X$. Block $k$, $k \in \llbracket 8 \rrbracket$, extracts temporal features and outputs a tensor $x_k \in \R^{(4 \times 2^k) \times C \times T / 2^k}$. Block 9-a performs the classification of any potential event $i \in \llbracket N_d \rrbracket$ and Block 9-b predicts the encoded center and duration of this event. Each convolution is followed by a zero padding layer and batch normalization. Note that the batch dimension is omitted. We have $\rho = N_d / (T / 2^8)$.}
\label{tab:network_architecture}

\end{table*}

%% file: 3_experiments.tex
\section{Experiments}

\paragraph*{Data}
\color{black}The experiments were performed on MASS SS2~\cite{OReilly2014}: 19 records from 19 subjects (11 females, 8 males, $\sim$ 23.6 $\pm$ 3.7 years old), sampled at $256$\,Hz\color{black}. The spindles have been scored by expert E1 (resp. E2) over $19$ records (resp. $15$) using different guidelines~\cite{Lajnef2017} \color{black}resulting in $\sim550$ (resp. $\sim1100$) scored spindles per record.  For records scored by both E1 and E2 $\sim 500$ spindles per record exhibit IoU $> 0$ (Gaussian-like distribution, pic at 0.6). The $15$ records annotated by E1 and E2 were used for spindles detection benchmark. \color{black} For K-complex detection, and joint spindle and K-complex detection, the $19$ records scored by E1 were used.

\paragraph*{Cross validation}
A $5$ split cross validation was used. A split stands for $10$ training, $2$ validation and $3$ (resp. $4$) testing records for spindle detection (resp. K-complex and joint spindle and K-complex detection).

\paragraph*{Metrics}
\emph{By event metrics}~\cite{Warby2014} were used to quantify the detection and localization performances of detectors. They rely on an $\mathrm{IoU}$ criterion: for a given $\delta > 0$, a predicted event was considered as a true positive if it exhibited an $\mathrm{IoU} \geq \delta$ with a true event otherwise it was considered as a false positive. The numbers of positives and true positives were evaluated to compute the precision, the recall and the F1 scores of detectors. Evaluation was performed on entire testing records. Performances reported were averaged over testing records.

\paragraph*{Baselines}
For spindles detection, $8$ baselines were benchmarked: \emph{Ferrarelli et al. 2007}~\cite{Ferrarelli2007}, \emph{M{\"{o}}lle et al. 2011}~\cite{Moelle2011}, \emph{Nir et al. 2011}~\cite{Nir2011}, \emph{Wamsley et al. 2012}~\cite{Wamsley2012}, \emph{Ray et al. 2015}~\cite{Ray2015} (Python package~\url{https://github.com/wonambi-python/wonambi}), \emph{Wendt et al. 2012}~\cite{Wendt2012}, \emph{Parekh et al. 2017}~\cite{Parekh2017} and \emph{Lajnef et al. 2017}~\cite{Lajnef2017}. We used the implementations by the authors of~\cite{Warby2014, Parekh2017, Lajnef2017}. For K-complex detection, \emph{Lajnef et al. 2017}~\cite{Lajnef2017} was compared.

Signal from C3 channel was used by all the baselines except \emph{Parekh et al. 2017} which used: F3, F4, Fz, C3, C4, Cz channels. Hyper-parameters of \emph{Parekh et al. 2017} and \emph{Lajnef et al. 2017} were selected in the ranges provided by the authors of~\cite{Parekh2017, Lajnef2017} by grid search on the validation data.

\paragraph*{Proposed approach}
The proposed approach was benchmarked on signal from channel C3. The network was provided with $20$\,s samples $x \in \R^{C \times T}$ with $C=1$ and $T = 5120$ ($256\,Hz$ sampling). A normalization was applied to $x$: centering and standardization by dividing each centered signal by its standard deviation computed on the full record.

The approach was implemented with PyTorch library~\cite{paszke2017automatic}. Minimizing~\eqref{eq:training_objective} was achieved with stochastic gradient descent using a learning rate $\mathrm{lr} = 10^{-3}$, a momentum $\mu = 0.9$ and a batch size of $32$. $100$ training epochs were considered. Each sample was randomly selected to contain at least a true event. When a true event was partially included in a sample, its label $l$ was set to $0$ if less than $50\%$ of this event was part of that sample. Early stopping was used to stop the training process when no improvement was observed on the loss evaluated on validation data after $5$ consecutive epochs.

Matching hyper-parameter $\eta$ was fixed to $\eta = 0.5$. Default event hyper-parameters were fixed to: $\rho=4$, $N_d = 80$, $t_i^d=256$. The resulting set $D(x) = \{( 64 \cdot i - 32, 256): i \in \llbracket N_d \rrbracket \}$ seemed a reasonable choice given the fact that both spindles and K-complexes have $\sim 1$\,s duration. A potential event $i$ was considered as a positive event of label $l \in \L$ if $\hat{\pi}_i^l \geq \theta^l$. Hyper parameter $\theta^l$ was selected by grid search over the validation data.
%
%

\begin{figure*}[ht!]
\centering
\includegraphics[width=0.92\linewidth]{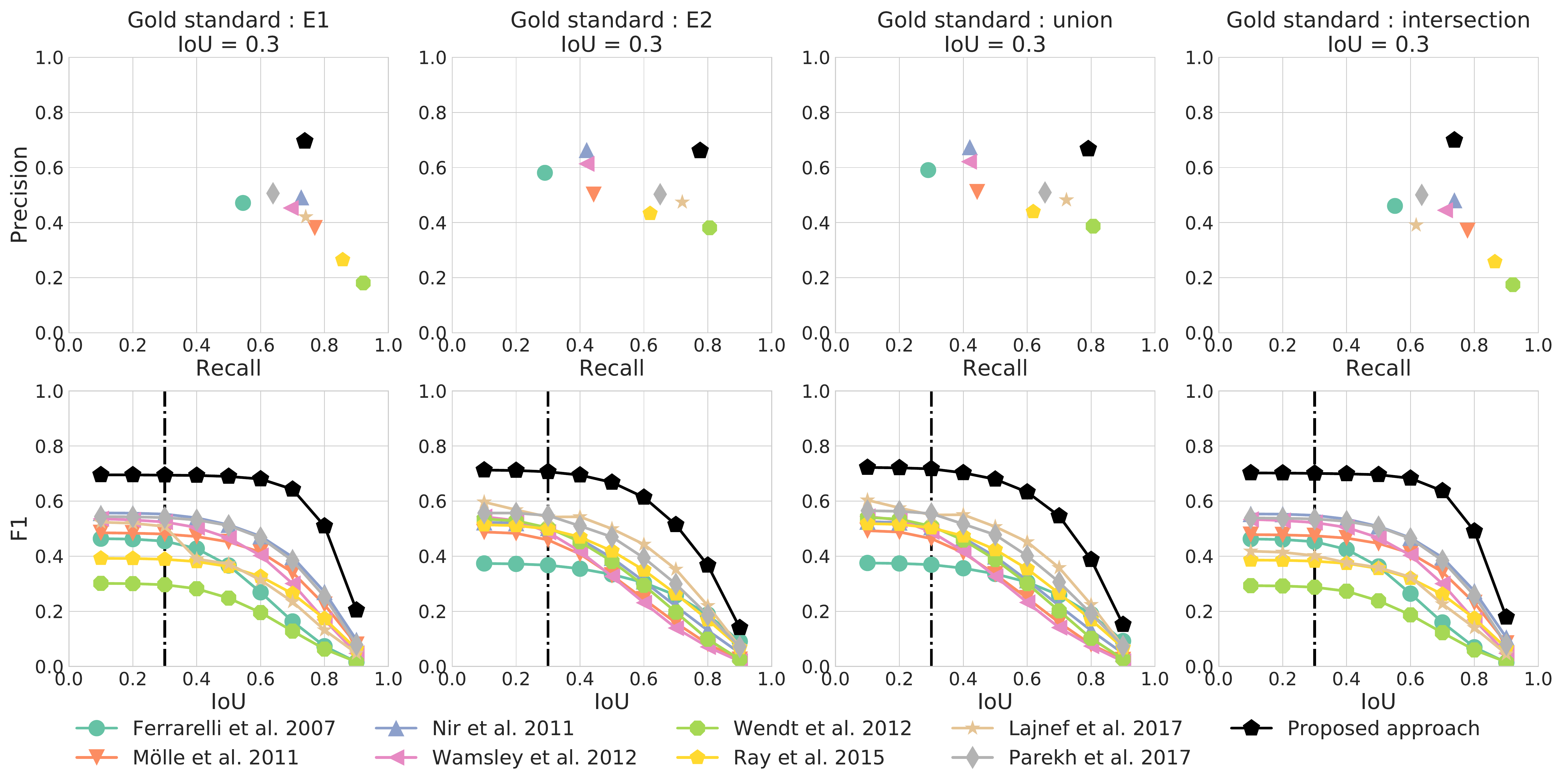}
\caption{Spindle detection: benchmark with respect to 4 gold standards: the proposed approach outperforms the baselines. First row: averaged precision / recall at $\mathrm{IoU} = 0.3$. Second row: F1 score as a function of IoU}
\label{fig:by_event_metrics}
\end{figure*}

\paragraph*{Spindles}
Detectors were compared to $4$ gold standards: events scored by E1, E2, the union and the intersection of events scored by both E1 and E2, see Figure~\ref{fig:by_event_metrics}.

First, the proposed approach seems to detect the occurrence of spindles without any supplementary information regarding the sleep stages but also to localize the spindles accurately. Indeed, the proposed approach outperforms the baselines in terms of precision / recall at $\mathrm{IoU} = 0.3$ and exhibits an higher F1 than the baselines for any $\mathrm{IoU}$. 
Second, the proposed approach seems to take into account any considered gold standard. Indeed, it exhibits stable performances over the gold standards contrarily to most of the baselines, except \emph{Parekh et al. 2017} and \emph{Lajnef et al. 2017}.

\begin{figure}[ht!]
\centering
\includegraphics[width=0.95\linewidth]{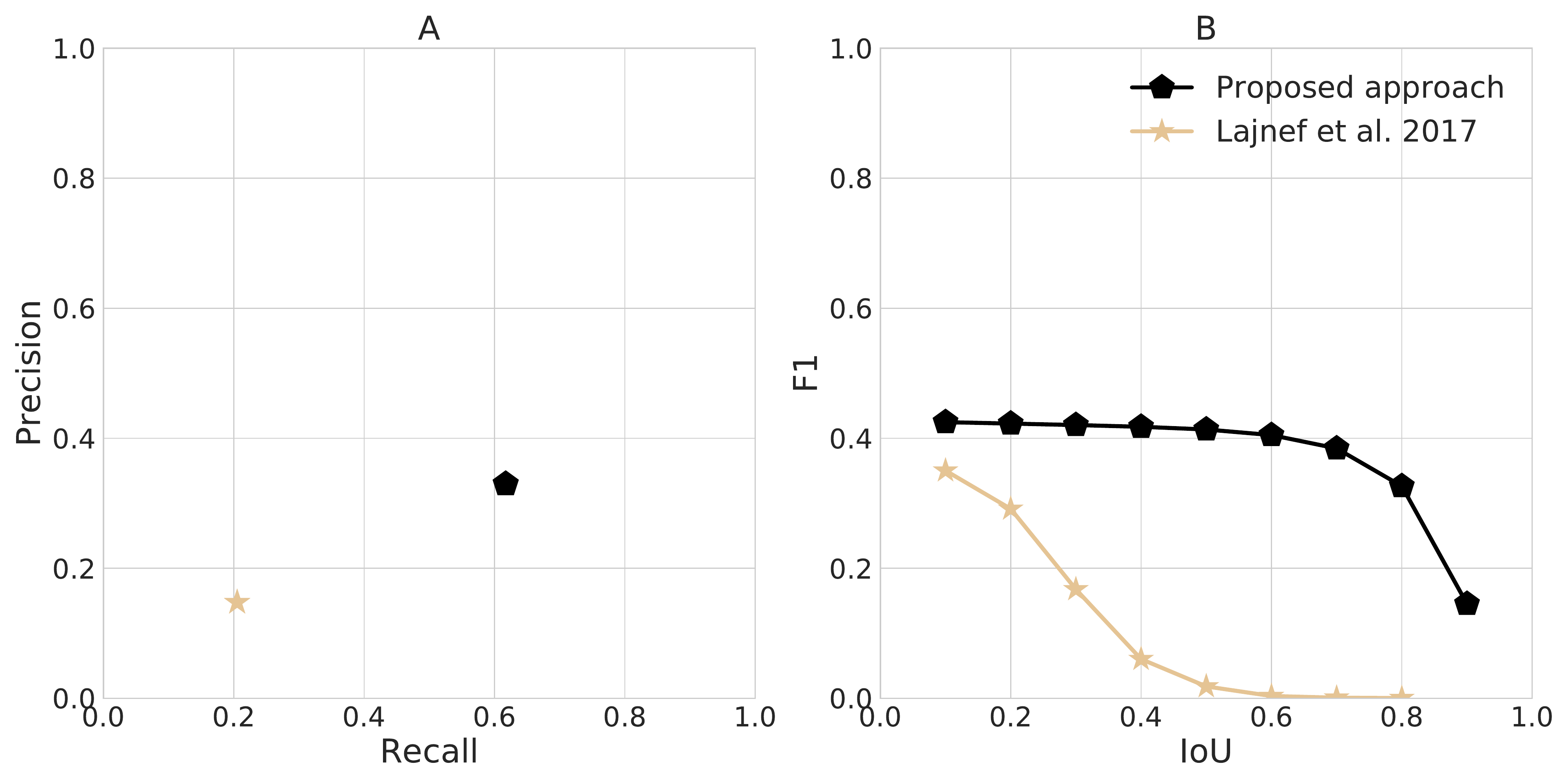}
\caption{K complex detection. A: precision / recall at $\mathrm{IoU} = 0.3$. B: F1 score as a function of IoU.}
\label{fig:by_event_k_complex}
\end{figure}

\begin{figure}[ht!]
\centering
\includegraphics[width=0.95\linewidth]{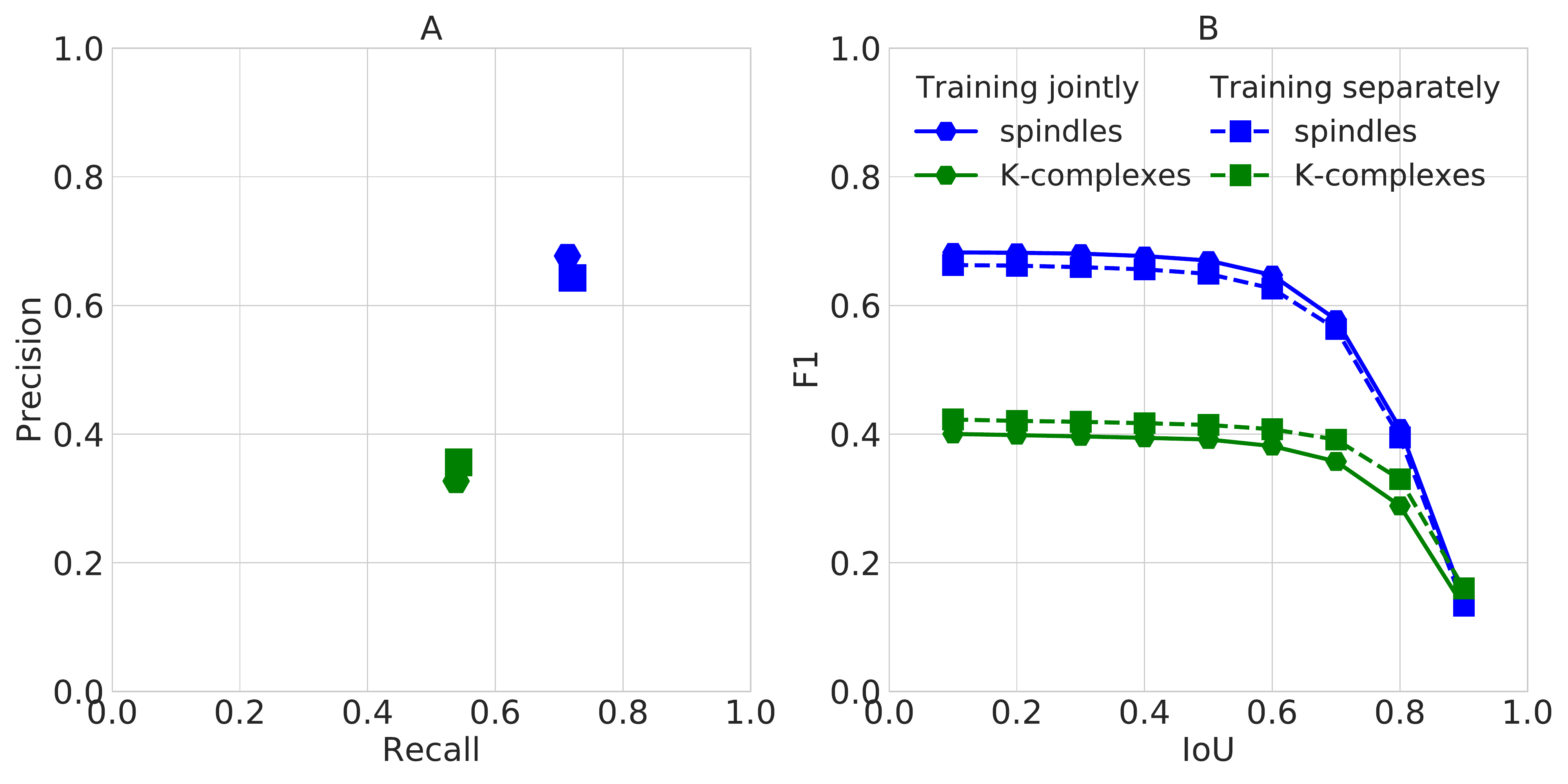}
\caption{\color{black}Detecting spindles and K-complexes jointly or separately leads to similar performances. A: Precision / Recall of detectors at IoU = 0.3. B: F1 scores as a function of IoU.\color{black}}
\label{fig:joined_vs_separate_training}
\end{figure}

\paragraph*{K-complexes}
Performances are reported in Figure~\ref{fig:by_event_k_complex}.
The proposed approach seems to outperform the baseline in terms of precision / recall at $\mathrm{IoU} = 0.3$ and exhibits a higher F1 score than the compared baseline at any IoU.
%
%
%
\paragraph*{Detecting events jointly or separately}
The proposed approach was trained to detect both spindle and K-complex jointly and separately. Performances are reported in Figure~\ref{fig:joined_vs_separate_training}. Same performances are obtained when the method is trained to detect spindles and K-complexes jointly or separately.

%% file: 4_discussion.tex
\section{Discussion}
The proposed approach builds on deep learning to learn a feature representation relevant for detecting any type of event. Surprisingly enough, the approach handles well the task of detecting spindles and K-complexes using \emph{only} $10$ training and $2$ validation records. We performed additional experiments (not shown) to vary the number of training records from $1$ to $10$: the method works when only $1$ training record is available. This might be due to random sampling which performs a kind of data augmentation.

%
%
As the proposed approach can handle multiple channels, additional experiments on spindles / K-complex detection were run using multiple channels: F3, F4, C3, C4, O1, O2. This did not result in any significant gain of performance on the used dataset (not shown). The method can also handle multiple modalities, electromyography (EMG),  electrooculography (EOG) or breathing, and mutiple default event scales at the same time, a property that was not explored in this study but that may be critical for detecting other types of events. This will be addressed in future studies.


%
%
The proposed approach seems to perform quite well with respect to different gold standards. Yet it remains to study how the method performs compared to the inter-scorer agreement~\cite{Warby2014}. This shall be also addressed in future works.

%% file: 5_conclusion.tex
\section{Conclusion}
This paper introduces a new deep learning architecture that can perform event detection of any type over an entire night of EEG recording. The proposed approach learns by back-propagation to build a feature representation from the raw input signal(s), and to predict both locations, durations and types of events. Numerical experiments on spindles and K-complexes detection demonstrate that the proposed approach outperforms previously published detection methods. A major advantage of the proposed approach is that it can detect jointly multiple types of events.